\documentclass[fleqn,oneside]{article}
\usepackage{espcrc2}

\usepackage{graphicx}
\usepackage{epsfig}

\newcommand{\AmS}{{\protect\the\textfont2
  A\kern-.1667em\lower.5ex\hbox{M}\kern-.125emS}}

\title{Properties of small HTSC mesa structures: common problems of interlayer tunneling}

\author{V.M.Krasnov\thanks{{\it E-mail address}:
krasnov@fy.chalmers.se}
\newline
\addressmark[]{Department of Microelectronics and Nanoscience, Chalmers
University of Technology, S-41296 G\"oteborg, Sweden}
 }

\begin{document}

\begin{abstract}
I analyze common problems of interlayer tunneling in Bi-2212 mesa
structures, such as self-heating and nonuniformity of junctions.
Numerical simulations have shown that self-heating does not mask
the temperature dependence of the superconducting gap. Major
problems can be avoided by decreasing mesa sizes.\\


\end{abstract}

\maketitle

\section{Introduction}

Layered high-$T_c$ superconductors (HTSC) represent natural stacks
of atomic scale intrinsic Josephson junctions (IJJ's). Indeed,
interlayer spacing of 15.5 $\AA$ was estimated from a periodic
Fraunhofer modulation of Fiske steps in Bi2212 mesas \cite{Fiske}.
IJJ's with their record-large $I_c R_n$ values $\sim 10 mV$ are
attractive candidates for cryoelectronics. Interlayer tunneling
spectroscopy has become a powerful tool for fundamental studies of
HTSC \cite{KrasnovT,KrasnovH}. However, interlayer tunneling has
several problems, such as defects, nonuniformity of IJJ's in the
mesa and self-heating or non-equilibrium quasiparticle injection
during transport measurements.

Here I analyze typical consequences of those problems. Numerical
simulations have shown that {\it self-heating does not mask
temperature dependence of the superconducting gap} close to $T_c$.
Major problems could be effectively avoided by decreasing the
in-plane size of mesas due to a proportional decrease of defects
and self-heating.

Fig. 1 shows experimental current-voltage characteristics (IVC's),
normalized per IJJ, (three probe configuration) for an
intercalated \cite{KrasnovH} HgBr$_2$-Bi2212 mesa $10 \times 20 \
\mu m^2$ at 3.9K $\leq T \leq$ 80K $>T_c \simeq 77 K$. IVC's
reveal a behavior typical for SIS tunnel junctions. The knee in
the IVC's represents the sum-gap voltage $V_s=2\Delta/e$, which
vanishes with approaching $T_c$ \cite{KrasnovT} and is suppressed
by magnetic field \cite{KrasnovH}. Temperature dependence of the
superconducting gap, $\Delta$, is shown in inset together with the
BCS curve.

\section{Self-heating}

An analytic solution of the self-heating problem for a circular
mesa yields \cite{Heating}:

\begin{equation}
T-T_0 = \frac{\pi q a}{4 \kappa_{ab}}, \label{Eq.1}
\end{equation}

where $T_0$ is the base temperature, $a$ is the radius of the
mesa, $q$ is the dissipated power density and $\kappa_{ab}$ is the
in-plane thermal conductivity. For Bi2212, self-heating may become
large at low $T$ due to a small $\kappa_{ab}$. From Fig. 1 it is
seen that there is back-bending of IVC's at the sum-gap voltage at
low $T$, due to self-heating and/or nonequilibrium quasiparticle
injection.

\begin{figure}[t]
\begin{minipage}{0.45\textwidth}
\epsfxsize=0.73\hsize \centerline{ \epsfbox{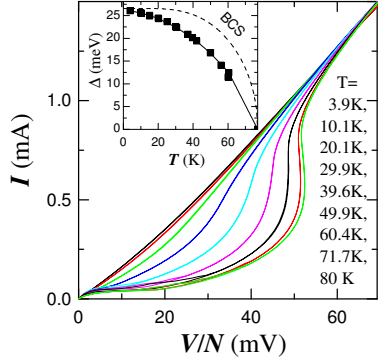} }
\end{minipage}
\vspace{-24pt} \caption{Experimental IVC's of a HgBr$_2$-Bi2212
mesa at 3.9K $\leq T \leq$ 80K. Inset shows the temperature
dependence of the superconducting gap.} \label{autonum1}
\end{figure}

Fig. 2 shows results of a numerical simulation of self-heating in
a circular mesa $a=5 \mu m$, containing $N=10$ IJJ's with
parameters typical for optimally doped Bi2212 \cite{KrasnovT}
($\rho_c = 30 \Omega cm$, $\Delta (T=0) = 30$ meV, $\kappa_{ab}(T)
\simeq 10^{-3} T$ (K) W cm$^{-1}$K$^{-1}$, $T_c = 95$ K).
Simulations were made in the following manner: First, a set of
IVC's($T$) without self-heating was calculated, see Fig. 2 a),
using standard SIS tunneling equations and a model $\Delta(T)$
dependence, shown by the solid line in Fig. 2 d). Calculations
were made for d-wave superconductors with coherent, directional
\cite{Millis} tunneling, providing a qualitative resemblance to
experimental data in Fig. 1. Next, the current was swept for a
fixed $T_0$ and equilibrium $V$ and $T$ were obtained by
numerically solving Eq.(1). Resulting IVC's for several $T_0$ are
shown in Fig. 2 b). It is seen that self-heating causes
back-bending of the IVC's, due to an increase of the effective
mesa temperature, as shown in Fig. 2 c).

The overheating is strongest at low $T$ both because of small
$\kappa$ and large $\Delta$. The "measured" $\Delta(T)$, obtained
from the maximum $dI/dV$ for increasing current, is shown in Fig.
2 d). It is smaller than the virgin $\Delta(T)$, however retains
the general shape. Indeed, at low $T$, despite large self-heating,
the variation of $\Delta$ is small due to a flat $\Delta (T)$
dependence, while close to $T_c$, self-heating vanishes as $q
\propto \Delta^2 \rightarrow 0$. Similar arguments apply for
nonequilibrium quasiparticle injection ($I \propto \Delta
\rightarrow 0$ at $T_c$).

\begin{figure}[t]
\begin{minipage}{0.49\textwidth}
\epsfxsize=.99\hsize \leftline{ \epsfbox{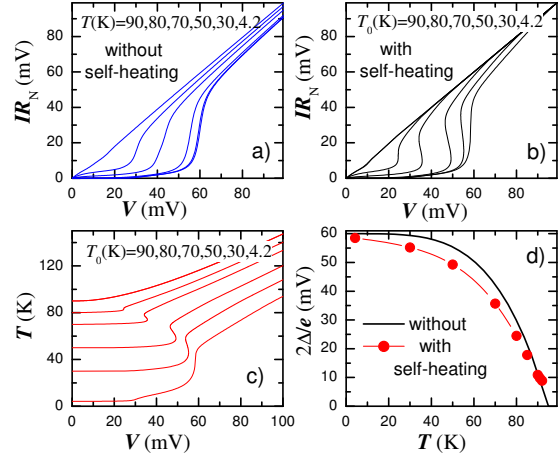} }
\end{minipage}
\vspace{-24pt} \caption{Numerical simulation of self-heating in a
Bi2212 mesa: a) and b) IVC's without and with self-heating at
different $T_0$; c) Temperature of the mesa along the IVC's from
b); d) Virgin (solid line) and measured (symbols) $\Delta(T)$. }
\label{autonum2}
\end{figure}

\section{Non-uniform junctions}

Fig. 3 a) shows quasiparticle branches in IVC's, (four probe
configuration), for a Bi2212 mesa at $T=4.2$K. Each branch
represents switching of an additional IJJ into a resistive state.
If all IJJ's in the mesa were identical, the multi-branch pattern
would have perfect periodicity \cite{KrasnovT}, while in Fig. 3 a)
a "ghost" branch "s" appears leading to a doubling of the branch
structure. The total set of branches corresponds to four identical
"w" junctions and one "s" junction. Junction "s" has a smaller
$V(I)$ and may either have smaller $\Delta$ (the IVC is squeezed
along $V-$axis) or have larger critical current, $I_c$ (the IVC is
stretched along $I-$axis). The latter assumption provides
excellent fit to the whole branch set if $I_c(s) \simeq 2.1
I_c(w)$, as shown by dashed lines in Fig. 3 a) for 1s, 1w and
1s+1w branches.

\begin{figure}[t]
\begin{minipage}{0.48\textwidth}
\epsfxsize=.99\hsize \leftline{ \epsfbox{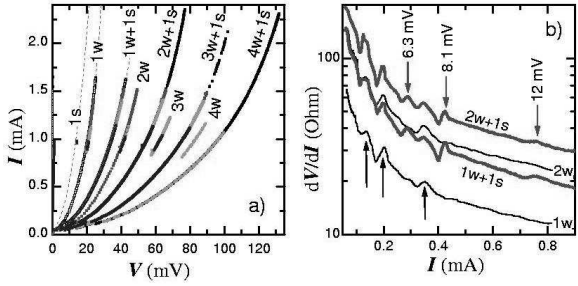} }
\end{minipage}
\vspace{-24pt} \caption{a) Quasiparticle IVC's of a Bi2212 mesa
with a "ghost" branch "s". Dashed lines represent a fit assuming
that "s" has 2.1 times larger area than "w". b) Sub-gap phonon
structures for different branches. In "s" the same phonon
structures appear at $\sim$ 2.1 times larger current.}
\label{autonum3}
\end{figure}

A decisive argument is provided from the analysis of sub-gap
phonon structures \cite{Schlen} in IVC's, shown in Fig. 3 b). For
the 1w branch, three well recognizable phonon structures at $\sim$
6.3, 8.1, and 12 mV, are marked in Fig. 3 b). Those features are
repeated at exactly the same current for the 2w branch, indicating
excellent uniformity of w-junctions. However, when the "s" branch
is involved, three additional maxima in $dV/dI$ appear. Those
features correspond to exactly the same phonon energies but occur
at $\sim$ 2.1 times larger currents. Therefore, junction "s" has
2.1 times larger $I_c$ but the same $\Delta$ as "w". Most probably
this is an underetched bottom junction with a larger area (note
that in a four probe configuration the top junction is not at the
surface of the crystal).

Fig. 4 a) shows experimental $dI/dV (V)$ curves (three probe
configuration) for two Bi2212 mesas on the same single crystal.
Both mesas contain $N=7$ IJJ's. It is seen that one mesa shows a
single sharp peak in conductance, while the other reveals several
maxima with smaller amplitudes.

Fig. 4 b) presents a numerical simulation of IVC's of a  mesa with
nonuniform IJJ's. The solid line represents the IVC for a d-wave
superconductor with coherent tunneling (without directionality cf.
with Fig. 1 a)). The corresponding tunneling conductance curve has
a sharp maximum at the maximum sum-gap voltage (in contrast to
usually considered incoherent d-wave tunneling). Symbols represent
the IVC per junction for a mesa with $N=3$, where the bottom IJJ
has 20\% larger area and $I_c$, the surface IJJ has 20\% smaller
$\Delta$ and $I_c$ than the middle IJJ (solid line). The total IVC
is strongly smeared and the corresponding $dI/dV (V)$ curve shows
three small maxima, in qualitative agreement with Fig. 4. a.

\begin{figure}[t]
\begin{minipage}{0.48\textwidth}
\epsfxsize=.99\hsize \leftline{ \epsfbox{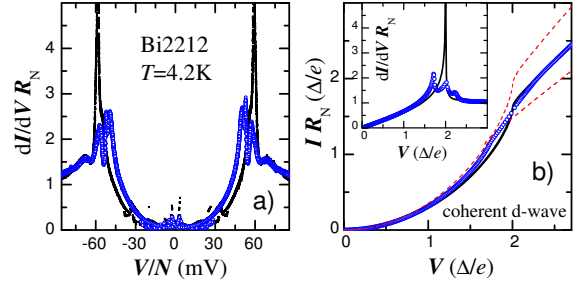} }
\end{minipage}
\vspace{-24pt} \caption{a) Experimental $dI/dV (V)$ curves for two
Bi2212 mesas from the same crystal. b) Simulated IVC's for a
single junction (solid line) and for a mesa with three slightly
different junctions (symbols). Corresponding $dI/dV$ curves are
shown in the inset.} \label{autonum}
\end{figure}

\section{Conclusions}

I have studied the influence of self-heating and non-uniformity of
junctions for interlayer tunneling in HTSC mesas. Numerical
analysis shows that {\it interlayer tunneling provides reliable
information about the superconducting gap even in the presence of
self-heating}. This is important in view of the highly debated
issue of whether $\Delta(T) \rightarrow 0$ at $T_c$
\cite{KrasnovT} or not \cite{Fisher}. Apparently, self-heating can
not explain the observed strong drop of $\Delta(T \rightarrow
T_c)$, if $\Delta$ were completely $T-$independent.

Non-uniformity of IJJ's deteriorates interlayer tunneling
characteristics. The only way of distinguishing genuine and
defect-induced properties is to check the reproducibility. E.g. in
case of Fig.4, similar single peak curves were observed for $\sim$
80\% of mesas, while the multi-peak structure (when present) was
different for each mesa. By decreasing the in-plane size and the
number of junctions in the mesa, major problems can be effectively
avoided due to a proportional decrease of defects and
self-heating.

\end{document}